\documentclass[aps,prl,showpacs,twocolumn,groupedaddress]{revtex4}
\newcommand \beq{\begin{eqnarray}}
\newcommand \eeq{\end{eqnarray}}
\def\simge{\mathrel{%
       \rlap{\raise 0.511ex \hbox{$>$}}{\lower 0.511ex \hbox{$\sim$}}}}
\def\simle{\mathrel{
       \rlap{\raise 0.511ex \hbox{$<$}}{\lower 0.511ex \hbox{$\sim$}}}}
\usepackage{graphics}
\usepackage[dvips]{graphicx}
\usepackage{graphicx}
\usepackage{color}

\newcommand{\rr}{\mathbf{r}}
\newcommand \la{\raisebox{-.5ex}{$\stackrel{<}{\sim}$}}

\begin{document}
\title{Coherence and clock shifts in ultracold Fermi gases with resonant
  interactions}
\author{Gordon Baym,$^{a,b}$ C.\ J.\ Pethick,$^{b,c}$
    Zhenhua Yu,$^a$ and  Martin W. Zwierlein$^{d,e}$}
\affiliation{\mbox{$^a$Department of Physics, University of Illinois, 1110
  W. Green Street, Urbana, IL 61801} \\
\mbox{$^b$The Niels Bohr Institute, Blegdamsvej 17, DK-2100 Copenhagen \O,
 Denmark}\\
\mbox{$^c$NORDITA, Roslagstullsbacken 23, SE-10691 Stockholm, Sweden }\\
\mbox{$^d$MIT--Harvard Center for Ultracold Atoms, Research Laboratory for
Electronics, Department of Physics, }\\{Massachusetts Institute of
Technology, Cambridge, MA 02139}\\
\mbox{$^e$Institut f\"ur Physik, Johannes Gutenberg-Universit\"at, 55099
Mainz, Germany}\\
}

\date{\today}

\begin{abstract}

    Using arguments based on sum rules, we derive a general result for the
average shifts of rf lines in Fermi gases in terms of interatomic interaction
strengths and two-particle correlation functions.  We show that, near an
interaction resonance, shifts vary inversely with the atomic scattering length,
rather than linearly as in dilute gases, thus accounting for the experimental
observation that clock shifts remain finite at Feshbach resonances.


\end{abstract}

\maketitle

    Interatomic interactions limit the accuracy of atomic clocks, causing
density-dependent {\it clock shifts\,} in radio frequency (rf) transitions.
Similarly, such shifts play an important role in probing correlations in
atomic gases~\cite{mit_two}, where, e.g., rf spectroscopy has been used to
detect the presence of molecules and provide evidence for pairing
gaps~\cite{regal,chin_rf,schunck}.  Surprisingly, experimentally observed
clock shifts become small when interactions are resonantly
enhanced~\cite{gupta,chin_rf}, a result we explain here, first by developing a
general theory of the average clock shift and then showing that in the
strongly interacting regime the shifts depend inversely on interatomic
scattering lengths.

    Single component spin-polarized Fermi gases do not experience clock
shifts, since the rf coupling preserves polarization, forbidding s-wave
interactions ~\cite{gupta}.  In addition, in mixtures of interacting fermions
in two states $|1\rangle$ and $|2\rangle$, e.g., the lowest two hyperfine
states of $^6$Li, clock shifts are absent for transitions $|1\rangle \to
|2\rangle$, since the interaction energy is invariant under the rf field
~\cite{mit_two}.  Interactions in two-state mixtures can be probed, rather, by
driving transitions to an initially empty state, e.g., in $^6$Li, from
$|2\rangle$ to the next hyperfine state $|3\rangle$.

    Rf transitions in atoms are usually described in terms of coherent
evolution of a two-level system undergoing Rabi oscillations, represented by
rotations of an equivalent pseudospin on the Bloch sphere.  In the present
problem, the rf field rotates an atom in state $|2\rangle$ into a coherent
superposition, $|2\rangle \to |\beta\rangle = \cos \theta |2\rangle +
e^{-i \phi}\sin \theta |3\rangle$.  (The angles $\phi$ and $\theta$ depend
on time, the rf pulse power, and its detuning from resonance.)  To the extent
that interatomic interactions only shift the energy levels, but do not broaden
the lines, the Bloch sphere picture is valid.  However, strong interactions
lead to an incoherent and irreversible evolution, and in the long-time, weak
pulse regime the probability to find the system in a particular final state is
given rather by Fermi's Golden Rule.

    We base our discussion on linear response theory (in Ref.~\cite{gupta}
fewer than 30\% of the atoms are transferred by the rf pulse, with comparable
numbers in Ref.~\cite{chin_rf}).  We consider a spatially uniform Fermi gas
with three internal states, and Hamiltonian $H = H_0+H_{\rm s}$, where $H_0
=\sum_{i=1}^3 \epsilon_iN_i$, $\epsilon_i$ is the hyperfine plus Zeeman energy
of level $|i\rangle$, and $N_i$ is the total number of atoms in state
$|i\rangle$.  The system Hamiltonian is ($\hbar=1$ throughout),
\begin{eqnarray}
 H_{\rm s}= \int d^3r \left(\frac{1}{2m} \nabla
 \psi_i^\dagger(\rr)\cdot\nabla\psi_i(\rr)\right)
 \nonumber\\
 +\sum_{i<j} \int d^3r d^3r' v_{ij}(\rr-\rr')
 \psi_i^\dagger(\rr)\psi_j^\dagger(\rr') \psi_j(\rr') \psi_i(\rr).
 \label{e:ham}
\end{eqnarray}
Here $v_{ij}(\rr - \rr')$ is the bare potential between two atoms in
$|i\rangle$ and $|j\rangle$ (not a low-energy effective interaction).

    The (essentially spatially uniform) rf field primarily couples the states
$|2\rangle$ and $|3\rangle$; other internal transitions are far off-resonance.
We therefore describe the rf coupling by $H_{\rm rf} = {\mathcal B(t)} Y$ with
the pseudospin-flip operator
\begin{equation}
 Y= i\int d^3r \left(\psi_3^\dagger(\rr)\psi_2(\rr)
 -\psi_2^\dagger(\rr)\psi_3(\rr)\right),
\label{e:Y}
\end{equation}
where ${\cal B}(t) = 2\omega_{\rm R}\cos\,\omega t$, with $\omega_{\rm R}$
the Rabi frequency.

    We assume an initial many-body state $|12\rangle$ of energy $E_{12}$,
containing atoms in states $|1\rangle$ and $|2\rangle$.  The transition rate
of atoms from state $|2\rangle$ to $|3\rangle$, $I(\omega)$, is
\begin{equation}
  I(\omega) = 2\pi \omega_{\rm R}^2 \sum_f |\langle f|Y|12\rangle|^2
\delta(\omega
+ E_{12}
  - E_f) \equiv 2\omega_{\rm R}^2\chi''(\omega),
\label{e:Fermi}
\end{equation}
where the sum runs over all eigenstates $|f\rangle$.  The mean frequency
for the transition,
\begin{equation} {\bar\omega} \equiv \frac{\int_{-\infty}^\infty d\omega
  \omega \chi''(\omega)}{\int_{-\infty}^\infty d\omega \chi''(\omega)}
   \equiv \omega_0+\Omega_{\rm c},
\end{equation}
where $\omega_0 =\epsilon_3-\epsilon_2$, and $\Omega_{\rm c}$ is the
average clock shift.  Using Eq.~(\ref{e:Fermi}) we derive the f-sum rule for
the first moment of the spectrum,
\begin{eqnarray}
 \int_{-\infty}^\infty
  \frac{d\omega}{\pi} \omega \chi''(\omega) &=& \sum_f (E_f - E_{12}) |\langle
  f|Y|12\rangle|^2 \nonumber \\
  &=& \frac{1}{2}\langle 12|[[Y,H],Y]|12\rangle.
\end{eqnarray}
For state $|3\rangle$ initially unoccupied,
\begin{eqnarray}
\int^{\infty}_{-\infty}\frac{d\omega}{\pi}\chi''(\omega) \hspace{128pt}
\nonumber \\
 =\int d^3r d^3r'\langle 12|\psi_2^\dagger(\rr)\psi_3(\rr)
  \psi_3^\dagger(\rr')\psi_2(\rr')|12\rangle = N_2,
\end{eqnarray}
where $N_2$ is the number of atoms in state $|2\rangle$.  We thus derive the
simple result for the mean clock shift
\begin{equation}
  \Omega_{\rm c} = \frac{1}{2 N_2} \langle 12|[[Y,H_{\rm s}],Y]|12\rangle.
\label{hh}
\end{equation}
More generally, the expectation value becomes an ensemble average.

    The average shift is simply the change in energy per 2-atom required to
rotate all $|2\rangle$ atoms in state $|12\rangle$ infinitesimally into state
$|\beta\rangle$.  This rotation transforms $|12\rangle$ into a state
$|1\beta\rangle = e^{i\theta Y}|12\rangle$ of $|1\rangle$ and $\beta$
atoms, with the same spatial many-particle wave function as $|12\rangle$.  The
energy difference of the two states is thus $\delta E = \langle
1\beta|H|1\beta\rangle - \langle 12|H|12\rangle \to \frac{1}{2}\theta^2
\langle 12|[[Y,H],Y]|12\rangle$ to second order in $\theta$.  Since the number
of atoms $\delta N$ transferred from $|2\rangle$ to $|3\rangle$ is $N_2
\theta^2$, the change in system energy per atom is given by Eq.~(\ref{hh}),
a result valid both for superfluid and normal phases. The general expression for the clock shift in terms of the interactions is
\begin{equation}
  \Omega_{\rm c} = \frac{1}{n_2}\int d^3r[v_{13}(r)-v_{12}(r)]\langle
  \psi_1^\dagger(\rr)\psi_2^\dagger(0)
  \psi_2(0)\psi_1(\rr)\rangle,  \nonumber\\
 \label{clock}
\end{equation}
where $n_2=N_2/V$, $V$ is the system volume, and $\langle\ldots\rangle$ denotes an ensemble average.

    To calculate the clock shift explicitly, we assume for simplicity in the
following that the interatomic potentials are short range contact
interactions, each with a large momentum cutoff.  We also assume that $n_1=n_2
=n$.  The bare couplings, $\bar g_{ij}$, are related to the measured low
energy couplings $g_{ij} = 4\pi a_{ij}/m$, where $a_{ij}$ is the scattering
length, by $g_{ij}^{-1}= {\bar g_{ij}}^{-1} +\int_0^\Lambda d^3p/(2\pi)^3
(m/p^2) = {\bar g_{ij}}^{-1} + m\Lambda_{ij}/2\pi^2$, as one sees by solving
the usual t-matrix equation.  Thus $\bar g_{ij} = 4\pi/\left(m(a_{ij}^{-1}+
r_{0,ij}^{-1})\right)$, where $r_{0,ij} = \pi/2\Lambda_{ij}$.  Physically we
expect $r_{0,ij}$ to be of order the characteristic length $(C_6 m/m_e)^{1/4}
a_0$ for the interatomic potential, where $C_6$ is the strength of the van der
Waals interaction; in $^6$Li, $r_0 \simeq 63 a_0$.  For short range contact
interactions, Eq.~(\ref{clock}) becomes
\beq
  \Omega_{\rm c} = \frac{1}{n_2}({\bar g}_{13}-{\bar g}_{12})\langle
  \psi_1^\dagger(0)\psi_2^\dagger(0)
  \psi_2(0)\psi_1(0)\rangle.
\label{omegaccorr}
\eeq

    The correlation function is related to the free energy, $F$, of the
system by the thermodynamic identity
\beq
 \frac{\partial (F/V)}{\partial {\bar g}_{12}}= \frac{1}{V}\langle {\partial
    H}/{\partial {\bar g}_{12}}\rangle = \langle
  \psi_1^\dagger(0)\psi_2^\dagger(0)
  \psi_2(0)\psi_1(0)\rangle.\nonumber\\
\eeq
We now take $r_0$ to be independent of the states of the atoms.
From Eq.\ (\ref{omegaccorr}) we then find the mean shift:
\beq
   \Omega_{\rm c} =
    \left(\frac{\bar a_{13}}{\bar a_{12}}\right)
     \left(\frac{1}{g_{13}}-\frac{1}{g_{12}}\right) \frac{1}{n_2}
     \frac{\partial (F/V)}{\partial g_{12}^{-1}}.
   \label{shifts}
\eeq

 When the magnitude of the scattering lengths are $ \gg r_0$,
which for $^6$Li is always valid above $B$ = 600G, the factor $\bar
a_{13}/\bar a_{12}$ is $\simeq 1$; the shift then involves only the
renormalized interactions, depending inversely on the scattering lengths.
This result has the same qualitative behavior, in the regime when $k|a| \gg
1$, where $k$ is a typical particle momentum, as predictions based on a
mean-field shift calculated from the real part of the forward scattering
amplitude in a dilute gas in the absence of effective range contributions
\cite{gupta}.

    The derivative $\partial(F/V)/\partial g_{12}^{-1} <0$ behaves perfectly
smoothly at the resonance in the 12-channel, and at zero temperature its value
there is $\sim -\epsilon_{\rm F}^2$, where $\epsilon_{\rm F} = p_{\rm F}^2/2m$
is the free particle Fermi energy and $p_{\rm F}$ is the Fermi momentum.  (At
characteristic densities $\sim 2 \times 10^{13} $cm$^{-3}$, $p_{\rm
F}^{-1}\sim 2000 a_0$ \cite{gupta}.)  We assume that $n_1=n_2 =n$ from here
on.  From the Monte Carlo calculations in Ref.~\cite{astr} of the energy $E$
at zero temperature we estimate that $\partial(E/V)/\partial g_{12}^{-1}
\simeq - 0.50 \epsilon_{\rm F}^2$ at the 12-resonance at $B \simeq 834$ G.
Then $\Omega_{\rm c} = g_{13}^{-1}\partial (F/V)/\partial g_{12}^{-1} \sim
-p_{\rm F}/ma_{13}$; this result, valid for $a_{13}\gg r_0$, scales as
$\sqrt{\epsilon_{F}}$.  On the other hand, at a 13-resonance ($B \simeq 690$
G), $\Omega_{\rm c} = -1/(n g_{12}) \partial (F/V)/\partial g_{12}^{-1}$.  As
we see, the average clock shift remains finite, even when scattering is
resonant.

    Figure 1 shows the correlation function $\partial (E/V)/\partial
a_{12}^{-1}$ calculated from the data of Ref.~\cite{astr} as a function of
$B$ for equal populations at zero temperature.  In the BCS limit, the
correlation function tends to the Hartree-BCS result, $n^2 + \Delta^2/\bar
g_{12}^2$, where $\Delta$ is the gap, while in the BEC limit ($p_{\rm
F}a_{12}\ll 1$) the system is composed entirely of 12-molecules with binding
energy $1/ma_{12}^2$, and $\partial E/ \partial g_{12}^{-1} = -4\pi
n/m^2a_{12}$, diverging as $a_{12}\to 0$.  For small positive scattering
length in a gaseous atomic state limit the correlation function tends to the
Hartree value $n^2$, as on the BCS side.  The large drop in $\partial E/
\partial a_{12}^{-1}$ seen in Fig.~1 for $B$ below $\sim 700$ G reflects the
presence of 12-molecules in the initial state.

\begin{figure}[correlationfunction]
{\centering \includegraphics[width=8cm]{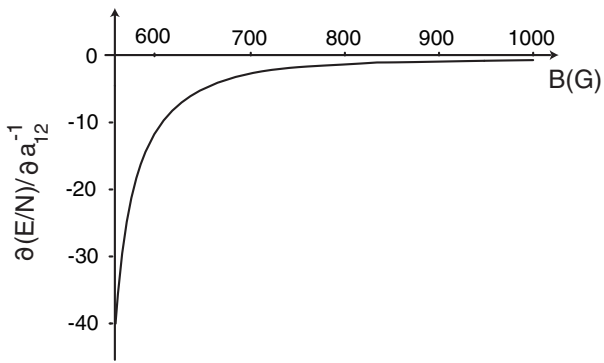}
\caption{Correlation function $\partial (E/N)/\partial a_{12}^{-1}$
in units of $p_{\rm F}/2m$ calculated from the data of Ref.\
\cite{astr}.}} \label{correlationfunction}
\end{figure}

    In the limit in which the bare couplings are small, $|{\bar a}| \ll r_0$, the
bare and renormalized couplings are equal, and the correlation function
becomes simply $n^2$; thus
\beq
  \Omega_{\rm c}=n(g_{13}-g_{12}),
\label{weakcoupling}
\eeq
the expected weak coupling result.  The intermediate regime in which $r_0
\ll |a| \ll p_{\rm F}^{-1}$ is more complicated since here short range
correlations as well as bound states in the appropriate channel play a role.
From Eq.~(\ref{shifts}) we find in this regime for an initial state without
12-molecules that
\beq
   \Omega_{\rm c} =
    \left(g_{12}-\frac{g_{12}^2}{g_{13}}\right)n =
   \left[\left(g_{13}-g_{12}\right)
  -\frac{\left(g_{13}-g_{12}\right)^2}{g_{13}}\right]n, \nonumber\\
   \label{intermediate}
\eeq
which differs from the weak coupling result (\ref{weakcoupling}) by a
factor ${g_{12}}/{g_{13}}$.  To derive this result, we note that the
correlation function in the absence of molecules is given by the Hartree
approximation, $n^2$, multiplied by the square of the Jastrow factor that
takes into account short-range correlations due to the interatomic
interaction.  At low energy, the Jastrow factor at the range of the forces is
essentially $1-a_{12}/r_0 = g_{12}/\bar g_{12}$, the ratio of the exact to the
non-interacting two particle wave function.  Thus $\langle
\psi_1^\dagger(0)\psi_2^\dagger(0) \psi_2(0)\psi_1(0)\rangle = (g_{12}/\bar
g_{12})^2 n^2$ \cite{integrate}.

    The difference between the result (\ref{intermediate}) and Eq.~(\ref{weakcoupling}) lies in
the distribution of spectral weight for transitions in the two cases.  The first
term $g_{13}-g_{12}$ in the last expression in Eq.~(\ref{intermediate}) is the
weak coupling result.

The second term however arises from short range
correlations, including, for $g_{13}>0$, 13-molecules. The squared matrix element for free-bound transitions is proportional to $(g_{13}-g_{12})^2$ \cite{julienne}, and the $1/g_{13}$ dependence reflects the fact that the binding energy of the molecule ($1/m a_{13}^2$) increases with decreasing $g_{13}$.
(Strongly bound 13-molecular states are not
described by the simple contact interaction model, but frequencies of
transitions to them in general lie outside the range explored in experiment.)
For $a_{13}<0$, there are no bound states, but higher-lying states with
kinetic energies of order $1/ma_{13}^2$, arising from short-range
correlations, extend the spectral weight out to frequencies of order $\sim
1/ma_{13}^2$, and contribute $\sim -1/a_{13}$ to the sum rule
\cite{lowenergy}.

    In the BEC limit, $p_{\rm F}a_{12}\ll 1$ the system is composed entirely
of molecules of binding energy $E_B = 1/ma_{12}^2$.  The energy density of the
system is just $E/V = -n/m a_{12}^2$, and from Eq.~(\ref{shifts}) we obtain
\beq
  \Omega_c = 2 E_B (1 - a_{12}/a_{13}),
  \label{molecules}
\eeq
as can also be calculated from Ref.~\cite{julienne}.

    In applying our results to experiment it is important to consider the
extent to which, for $a_{12}>0$, molecules are present in the initial state.
The experiment of Ref.~\cite{gupta} is carried out on a timescale short
compared with that to create diatomic molecules via three-body collisions.
Thus the correlation function entering the sum rule is that of the
molecule-free metastable state.  On the other hand, in the initial state in
the experiment of Ref.~\cite{chin_rf} molecular states were equilibrated, and
their effects must be included in the correlation function.  Furthermore
comparison with experiment requires that measurements be carried out over the
entire frequency range, on scales $\sim 1/ma_{13}^2$, where there are
significant contributions to the spectral weight.

    The solid line (a) and (b) in Fig.~2 shows the predicted clock shift for
equilibrated $^6$Li, in units of $\epsilon_{\rm F}/\hbar$ in the region $\sim$
600-1200 G at zero temperature.  Scattering lengths are taken from Ref.\
\cite{bartenstein}, and we assume $p_{\rm F}=1/2000 a_0^{-1}$.  The dashed-dot
line (c) shows the result (\ref{molecules}) for a non-interacting gas of
molecules.  The dashed part of curve (a) below the 12-resonance, the
difference of the full Monte-Carlo result (b) in the 12-molecular region, and
the two-body contribution (c), is a measure of the many-body contribution to
the shift below the 12-resonance.

\begin{figure}[htb]
{\centering \includegraphics[width=8.5cm]{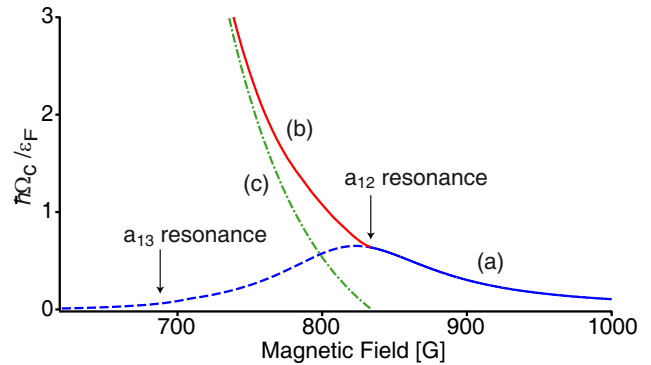}
\caption{Calculated clock shift for a gas of $^6$Li atoms at zero
temperature as a function of magnetic field.  See text for
details. }} \label{shift}
\end{figure}

    The correct picture of rf field excitation combines both coherent rotation
of the initial state with transitions that change the number of elementary
excitations of the system and thus lead to a width, $\Gamma$, of the spectrum
$I(\omega)$.  For times short compared to $\omega_{\rm R}^{-1}$ and
$\Gamma^{-1}$ the rf field indeed starts to rotate all $|2\rangle$ atoms
coherently into $|3\rangle$.  For a strong drive, with $\omega_{\rm R} \gg
\Gamma$, the many-body state undergoes Rabi oscillations, damped on a time
scale $\sim\Gamma^{-1}$.  The long-time behavior, when all oscillations have
damped out, is captured by the Golden Rule.

    Under an instantaneous rotation of $|2\rangle$ to $|\beta\rangle$, the
$|1\beta\rangle$ state produced has the same spatial wave function as
$|12\rangle$.  However the bare amplitude for $1,\beta \to 1,\beta$
scattering, $\bar g_{1\beta} =\cos^2\theta \bar g_{12} + \sin^2\theta
\bar g_{13}$, differs from the bare 12 amplitude by $\delta \bar g = \bar
g_{1\beta}-\bar g_{12} = \sin^2\theta (\bar g_{13}-\bar g_{12})$.  One can understand Eq.~(\ref{omegaccorr}) from this
point of view, by noting that since the spatial wave functions of
$|1\beta\rangle$ and $|12\rangle$ are identical, the difference of their
energies (and free energies at finite temperature) is, for small $\theta$,
$E_{1\beta}-E_{12}= \delta \bar g \langle \psi_1^\dagger(0)\psi_2^\dagger(0)
\psi_2(0)\psi_1(0)\rangle$.

    The rotated state $|1\beta\rangle$ is not in fact an energy eigenstate,
and thus the response has a width.  Under rotation the interaction energy
becomes
\beq
   H_{\rm int} = [\bar g_{12}\cos^2\theta
   + \bar g_{13}\sin^2\theta]
       \psi_1^\dagger\psi_\beta^\dagger \psi_\beta \psi_1
       \nonumber\\
       +[\bar g_{13}\cos^2\theta + \bar g_{12}\sin^2\theta]
       \psi_1^\dagger\psi_\alpha^\dagger \psi_\alpha \psi_1    \nonumber\\
     +\left(\bar g_{13} - \bar g_{12}\right)\sin\theta \cos\theta
       \,\,
     \psi_1^\dagger \left(\psi_\alpha^\dagger \psi_\beta + \psi_\beta^\dagger
     \psi_\alpha\right) \psi_1, \nonumber\\
   \label{fullint}
\eeq
where $|\alpha\rangle = \cos\theta |3\rangle - \sin\theta
|2\rangle $ is the superposition of single particle states $|2\rangle$ and
$|3\rangle$ orthogonal to $|\beta\rangle$.  The first term in
Eq.~(\ref{fullint}) is the original interaction with a modified coupling
constant, in terms of states $|1\rangle$ and $|\beta\rangle$, instead of
$|1\rangle$ and $|2\rangle$.  The second term acting on a rotated state
without $\alpha$ atoms present vanishes.  The final term, however, mixes the
two particles in states $|1\rangle$ and $|\beta\rangle$ into $|1\rangle$ and
$|\alpha\rangle$.  Thus while the external rf field coherently rotates the
initial state, the latter interaction decoheres the rotated state, leading to
a mixture of energy eigenstates and a response with nonzero width.  Even in
the absence of the final term, the rotated state is not an eigenstate of
$H_{\rm s}$, since the spatial wave function of an eigenstate with $\bar
g_{1\beta}$ differs from that with the original $\bar g_{12}$.

    The lowest order perturbative approach does not include collective
effects, e.g., screening of the interaction by the medium, and fails to
satisfy the sum rule in the presence of interactions.  As indicated in Ref.\
\cite{YB}, these effects must be included by summing chains of bubble
diagrams.  Such terms are especially important in the superfluid state, since
they take into account transitions to states in which the phase of the
condensate is altered globally.

    In summary, we have shown from sum-rule arguments that the average clock
shifts in strongly interacting Fermi gases are finite at Feshbach resonances.
The interaction model we have employed is admittedly crude, since it does not
allow for the many bound states that exist for real interatomic potentials.
Detailed calculations of the spectral distribution must be done
microscopically.  In addition to the self-energy corrections to propagators,
it is necessary to include vertex corrections, as pointed out in Ref.\
\cite{YB}.  Such effects are important even in a low density gas, where one
finds interference terms between 12-scattering and 13-scattering processes
\cite{koelman}.  These processes, which correspond to the Aslamazov-Larkin
diagrams for fluctuation-induced effects in superconductors \cite{AL}, will be
discussed in a future publication.  In addition, for detailed comparison with
experiment it is necessary to take into account the inhomogeneity of the
particle density.

\acknowledgements

    This research was supported in part by NSF Grants PHY03-55014,
PHY05-00914, and PHY07-01611.  We are grateful for the hospitality of the
ECT$^*$ in Trento (Italy) which enabled this collaboration, and thank E.
Demler, S. Giorgini, W. Ketterle, W. Phillips, W. Zwerger, and especially
H.T.C.~Stoof, for helpful input.


\begin{thebibliography}{20}
\bibitem{integrate} Integrating $\partial (E/V)/\partial g_{12}^{-1} =
-g_{12}^2n^2$ with respect to $\bar g$ we recover immediately the low-energy
result, $E_{\rm int}/V = g_{12}n^2$.
\end{thebibliography}
\end{document}